\documentclass[12pt,a4paper]{article}
\usepackage{amsmath}
\usepackage{amsfonts}
\usepackage{amssymb}
\usepackage{textcomp}
\usepackage[dvips]{graphicx}
\usepackage{epsfig}
\usepackage{bm}
\usepackage{dcolumn}
\usepackage{amsfonts}
\usepackage{color}
\usepackage[left=2cm,right=2cm,top=2cm,bottom=2cm]{geometry}
\usepackage[tableposition=top]{caption}
\usepackage{subcaption}
\DeclareCaptionLabelFormat{gostfigure}{Fig. #2}
\DeclareCaptionLabelSeparator{gost}{.~~}
\begin{document}
\title{\textbf{Late-time power-law stages of cosmological evolution in teleparallel gravity with nonminimal coupling}}
\author{Maria A. Skugoreva\footnote{masha-sk@mail.ru}\vspace*{3mm} \\
\small Kazan Federal University, Kremlevskaya 18, Kazan 420008, Russia}

\date{ \ }
\maketitle
\begin{abstract}

    We investigate the Universe evolution at late-time stages in models of teleparallel gravity with power-law nonminimal coupling and a decreasing power-law potential of the scalar field $\phi$. New asymptotic solutions are found analytically for these models in vacuum and with a perfect fluid. Applying numerical integration, we show that the cosmological evolution leads to these solutions for some region of the initial conditions, and these asymptotic regimes are stable with respect to homogeneous variations of the initial data. The physical sense of the results is discussed.  
\end{abstract}

\section{Introduction}

~~~~~~Final stages of cosmological evolution are studied widely both in General Relativity (GR) as in its modifications. It is interesting and urgent especially due to the need for an explanation of observational data \cite{observ1,observ2,observ3,observ4,observ5} (see also the review \cite{reobserv}) indicating the late-time accelerated expansion of the Universe. Realistic models should describe observations and be free of shortcomings. There are such problems as ``fine-tuning'' of initial conditions for realization of the late-time cosmic acceleration in models of $\Lambda$CDM \cite{Amendola} and quintessence \cite{Matarrese} based on GR, future singularities like a Big Rip, ``sudden'' and others (see, e.g., \cite{Tsujikawa,Bamba}). Cosmological evolution leads to a Big Rip in some models of modified gravity \cite{f(R)1,f(R)2,f(R)3,f(R)4,f(RBoxR)}. 

     There is an alternative formulation of GR, teleparallel gravity (Teleparallel Equivalent of General Relativity, TEGR \cite{JGPereira}). It is based, firstly, on Einstein's idea of absolute parallelism \cite{Einstein1,Einstein2}, that is on using of a field of orthonormal bases --- tetrads --- for tangent space-times and, secondly, TEGR applies the Weitzenb\"{o}ck connection \cite{Weitzenbock} instead Levi-Civita one, which leads to zero curvature and nonzero torsion. The TEGR Lagrangian contains the torsion scalar $T$, and equations of motion of this theory coincide exactly with those of GR \cite{TEGR1,TEGR2,Hayashi:1979qx,Maluf:2013gaa}. However, modifications of teleparallel gravity (for example, $f(T)$ theory \cite{Ferraro1,Ferraro2,Linder:2010py,Cai}, and scalar-torsion gravity \cite{Geng1,Geng2,Xu:2012jf,Wey,Otalora:2013tba,Geng3,Geng4,Sadjadi1,Kucukakca,Sadjadi2}) are not equivalents of similar modifications of GR, their difference consists in a term with the divergence of torsion in the Lagrangian (see \cite{Bahamonde}). It gives rise to different field equations and consequently to a new cosmological dynamics. Therefore, it is of interest to investigate modifications of teleparallel gravity. Scenarios with the late-time cosmic acceleration were already found, for example, in teleparallel gravity with nonminimal coupling of the form $\xi T\phi^2$ \cite{Geng1,Geng2,Xu:2012jf,Wey,Otalora:2013tba}. 
    
    In our recent papers \cite{Masha1, Masha2}, stable asymptotic solutions (attractors of the corresponding dynamical systems) were obtained in models of teleparallel gravity with a nonminimal coupling of the form $\xi T\phi^N$ for $N=2$ and a potential of the scalar field $V(\phi)=V_0\phi^n$. Those methods of dynamical system theory did not allow us to investigate cases of $N>2$, $\xi>0$ and ~~$n<0$. In the present work we shall find the final stages of the Universe evolution in such models which are not studied earlier. Units $\hbar=c=1$ will be used. 

     The structure of this paper is as following. In Section~\ref{II} we briefly present foundations of teleparallel gravity and write the basic equations for the considered models of scalar-torsion gravity. In Section~\ref{III} the analytic and numerical results are described. They are discussed in Section~\ref{IV}.    
      
\section{Basic equations}
\label{II}

~~~~Let us shortly present the foundations of teleperallel gravity and write the main equations for its modification with a nonminimally coupled scalar field.
    
    In teleparallel gravity, the dynamical variables are four linearly independent vectors, the tetrad $\mathbf{e}_A(x^{\mu})=e^{\mu}_A\partial_{\mu}$, where Greek indices are space-time ones, capital Latin indices are tangent space-time those. A tetrad forms an orthonormal basis in the tangent space at each point of space-time. Then the metric tensor is 
\begin{equation}
\label{1}
g_{\mu\nu}=\eta_{AB}e^{A}_{\mu}e^{B}_{\nu},
\end{equation}
where $\eta_{AB}={\rm diag}(1,-1,-1,-1)$ is the Minkowski metric tensor, $\eta_{AB}=\mathbf{e}_A\cdot\mathbf{e}_B$. The determinant consisting of tetrad components $e^A_{\mu}$ is $e=\text{det}(e^A_{\mu})=\sqrt{-g}$.

   In teleparallel gravity, the Wetzenb\"{o}ck connection is used:
\begin{equation}
\label{2}
\overset{\mathbf{w}}\Gamma^{\lambda}_{\mu\nu}\equiv e^{\lambda}_A\partial_{\mu}e^{A}_{\nu},
\end{equation} 
which leads to the curvature scalar $\overset{\mathbf{w}}R=0$, while the torsion tensor and the torsion scalar are
\begin{equation}
\label{3}
T^{\lambda}_{\mu\nu}\equiv\overset{\mathbf{w}}\Gamma^{\lambda}_{\nu\mu}-\overset{\mathbf{w}}\Gamma^{\lambda}_{\mu\nu}=e^{\lambda}_A(\partial_\mu e^A_{\nu}-\partial_\nu e^A_{\mu}),
\end{equation}
\begin{equation}
\label{4}
T\equiv\frac{1}{4}T^{\rho\mu\nu}T_{\rho\mu\nu}+\frac{1}{2}T^{\rho\mu\nu}T_{\nu\mu\rho}-T^{\rho}_{\rho\mu}T^{\nu\mu}_{\nu}.
\end{equation}
We ~~note ~~that ~~actually ~~the ~~torsion ~~scalar ~is ~a ~linear ~combination ~of ~three ~scalars:
$$
T=a T^{\rho\mu\nu}T_{\rho\mu\nu} +b T^{\rho\mu\nu}T_{\nu\mu\rho} +c T^\rho_{\rho\mu}T^{\nu\mu}_\nu. 
$$
Constant coefficients ~$a$, ~$b$, ~$c$~ are chosen so that the field equations of teleparallel gravity are equivalent to those of GR (see \cite{Einstein3}).

    The TEGR action has the form 
$$
S=\frac{1}{16\pi G}\int T\ e\ d^4x.
$$ 
If we add to it the nonminimally coupled scalar field with a potential and also matter, the action is
\begin{equation}
\label{5}
S=\frac{1}{2}\int \left[T\left( \frac{1}{K}+\xi B(\phi)\right) +\partial_{\mu}\phi\partial^{\mu}\phi-2V(\phi)\right] e\ d^4x+S_m, 
\end{equation}
where $\xi$ is the coupling constant, $B(\phi)$ is the nonminimal coupled function, $V(\phi)$ is the potential of the scalar field $\phi$, ~~$S_m$ is the matter action, and $K=8\pi G$.
\\

    A spatially flat Friedmann-Lema\^{i}tre-Robertson-Walker metric will be used:
\begin{equation}
\label{6}
ds^2=dt^2-a^2(t)\delta_{ij}{dx}^i {dx}^j,
\end{equation}
where $i,j=1,2,3$ are the spatial indices, ~~$t$ the cosmic time, $x^{i}$ the comoving spatial coordinates, and $a(t)$ the scale factor.

    The gravitational and scalar field equations are derived by varying the action (\ref{5}) with respect to the tetrad $e^A_\mu={\rm diag}(1,a(t), a(t), a(t))$, which corresponds to the chosen metric (\ref{6}), and with respect to the scalar field:
\begin{equation}
\label{7}
3H^2=K\left( \frac{{\dot{\phi}}^2}{2}+V(\phi)-3\xi H^2 B(\phi)+\rho\right) ,
\end{equation}
\begin{equation}
\label{8}
2\dot{H}=-K\left( {\dot{\phi}}^2+2\xi H \dot{\phi}B'(\phi)+2\xi\dot{H}B(\phi)+\rho(1+w)\right) ,
\end{equation}
\begin{equation}
\label{9}
\ddot{\phi}+3 H\dot{\phi}+3\xi H^2 B'(\phi)+V'(\phi)=0.
\end{equation}
Here $H(t)\equiv\frac{\dot a}{a}$ is the Hubble parameter, $\rho$ the energy density of matter, $p$ the pressure of matter, $p=w\rho$ the matter equation of state, $w\in[-1;1]$ is a constant, a dot denotes a time derivative, and a prime is a derivative with respect to the scalar field. The torsion scalar is $T=-6H^2$ in the tetrad chosen.

    It should be noted that generalizations of teleparallel gravity are not invariant under local Lorentz transformations (see, e.g., \cite{Sotiriou}), therefore the choice of a tetrad influences the form of the field equations. It was already shown for $f(T)$ theory of gravity in \cite{Krssak} that the flat FLRW tetrad in Cartesian coordinates has advantages. Therefore, we can expect that this tetrad will also be preferable in scalar-torsion gravity.

    Equations (\ref{7}), (\ref{8}) can be rewritten as
\begin{equation}
\label{10}
3H^2=K(\rho_{\phi}+\rho),
\end{equation}
\begin{equation}
\label{11}
2\dot H=-K(p_{\phi}+\rho_{\phi}+p+\rho),
\end{equation}
where 
\begin{equation}
\label{12}
\rho_{\phi}=\frac{{\dot{\phi}}^2}{2}+V(\phi)-3\xi H^2 B(\phi),
\end{equation}
\begin{equation}
\label{13}
p_{\phi}=\frac{{\dot{\phi}}^2}{2}-V(\phi)+2\xi H \dot{\phi}B'(\phi)+\xi(2\dot{H}+3H^2)B(\phi),
\end{equation}

    In this work we consider cosmological models with $B(\phi)=\phi^N$, $N>2$ even; ~~$V(\phi)=V_0\phi^n$, $n<0$ even; ~~$\xi>0$. The parameters $N$, $n$, $w$ are dimensionless.

\section{Asymptotic solutions and numerical analysis of their stability}
\label{III}
\subsection{The Vacuum Case}
\label{III.1}\
~~~~Several asymptotic power-law regimes (such that $\phi\to\infty$) have been found in cosmological models of telleparallel gravity with nonminimal coupling $\xi T\phi^N$ for $N=2$ and the potential of the form $V(\phi)=V_0\phi^n$ in our previous papers \cite{Masha1, Masha2}. As the asymptotic power-law behavior of the Hubble parameter $H(t)$ and the scalar field $\phi(t)$ is quite typical in such models, we could expect the emergence of similar solutions in the case $N>2$ as well. Cosmological models with $N>2$ have not been studied in \cite{Masha1, Masha2} using expansion-normalized variables since the corresponding dynamic system has a zero denominator for $N\neq2$, ~~$\phi\to\infty$.

    We want to check whether or not there is asymptotic regime of a power-law form of $H(t)$ and $\phi(t)$ in the models under study. Let us substitute the solution ~~$H(t)=H_0{(t-t_0)}^\alpha$, ~~$\phi(t)=\phi_0{(t-t_0)}^{\alpha\beta}$, ~~$\rho=0$, ~~where ~~$\alpha$, ~~$\beta$, ~~$H_0$, ~~${\phi}_0$, ~~$t_0$~~ are constants, to the system (\ref{7})-(\ref{9}), where ~~$B(\phi)=\phi^N$, ~~$V(\phi)=V_0\phi^n$.
\begin{equation}
\label{14}
\begin{array}{l}
3{H_0}^2{(t-t_0)}^{2\alpha}=K\left[ \frac{\alpha^2\beta^2}{2}{\phi_0}^2{(t-t_0)}^{2\alpha\beta-2}+V_0{\phi_0}^n{(t-t_0)}^{n\alpha\beta}-\right. \\
~~~~~~~~~~~~~~~~~~~~\left. -3\xi {H_0}^2{(t-t_0)}^{2\alpha}{\phi_0}^N{(t-t_0)}^{N\alpha\beta}\right] ,
\end{array}
\end{equation}
\begin{equation}
\label{15}
\begin{array}{l}
2\alpha H_0{(t-t_0)}^{\alpha-1}=-K\left[ \alpha^2\beta^2{\phi_0}^2{(t-t_0)}^{2\alpha\beta-2}+2\xi N H_0{(t-t_0)}^\alpha \alpha\beta{\phi_0}^N{(t-t_0)}^{N\alpha\beta-1}+\right. \\
~~~~~~~~~~~~~~~~~~~~~~~~\left. +2\xi \alpha H_0{(t-t_0)}^{\alpha-1}{\phi_0}^N{(t-t_0)}^{N\alpha\beta}\right] ,
\end{array}
\end{equation}
\begin{equation}
\label{16}
\begin{array}{l}
\alpha\beta(\alpha\beta-1)\phi_0{(t-t_0)}^{\alpha\beta-2}+3 H_0{(t-t_0)}^\alpha \alpha\beta\phi_0{(t-t_0)}^{\alpha\beta-1}+\\
+3\xi N{H_0}^2{(t-t_0)}^{2\alpha} {\phi_0}^{N-1}{(t-t_0)}^{(N-1)\alpha\beta}+n V_0{\phi_0}^{n-1}{(t-t_0)}^{(n-1)\alpha\beta}=0.
\end{array}
\end{equation}
We will consider two cases: \textbf{(i)} where the scalar potential is neglected and \textbf{(ii)} where the potential significantly affects the cosmological evolution. 
\\
\\
\textbf{i)} If ~~$\alpha<0$, ~~$\beta<0$, ~~$N>2$, ~~$n\alpha\beta<2\alpha\beta-2$~~ and
\begin{equation}
\label{17}
2\alpha\beta-2=N\alpha\beta+2\alpha \quad \Leftrightarrow\quad \alpha\beta=\frac{2(1+\alpha)}{2-N}>0 \quad \Rightarrow \quad \alpha<-1, 
\end{equation}
then ~~$2\alpha\beta-2<N\alpha\beta+\alpha-1$,~~ and neglecting smaller terms for $t\to+\infty$, we obtain
\begin{equation}
\label{18}
\begin{array}{l}
0=K\left[ \frac{\alpha^2\beta^2}{2}{\phi_0}^2{(t-t_0)}^{2\alpha\beta-2}-3\xi {H_0}^2{\phi_0}^N{(t-t_0)}^{N\alpha\beta+2\alpha}\right] ,
\end{array}
\end{equation}
\begin{equation}
\label{19}
\begin{array}{l}
0=-K\left[2\xi N H_0 \alpha\beta{\phi_0}^N{(t-t_0)}^{N\alpha\beta+\alpha-1}+2\xi \alpha H_0{\phi_0}^N{(t-t_0)}^{N\alpha\beta+\alpha-1}\right] ,
\end{array}{}
\end{equation}
\begin{equation}
\label{20}
\begin{array}{l}
\alpha\beta(\alpha\beta-1)\phi_0{(t-t_0)}^{\alpha\beta-2}+3\xi N{H_0}^2{\phi_0}^{N-1}{(t-t_0)}^{(N-1)\alpha\beta+2\alpha}=0.
\end{array}
\end{equation}
We obtain from (\ref{18}), (\ref{19}), (\ref{20}) using (\ref{17}):
\begin{equation}
\label{21}
\beta=-\frac{1}{N},~~~~\alpha=-{\frac{2 N}{N+2}},~~~~\alpha\beta=\frac{2}{N+2},
\end{equation}
\begin{equation}
\label{22}
{H_0}^2=\frac{2}{3\xi {(N+2)}^2}{\phi_0}^{2-N}.
\end{equation}
Recalling the initial assumption ~~$n\alpha\beta<2\alpha\beta-2$,~~ we get
\begin{equation}
\label{23}
n<-N.
\end{equation}
\\
\\
\textbf{ii)} If ~~$\alpha<0$, ~~$\beta<0$, ~~$N>2$~~ and ~~$n\alpha\beta=2\alpha\beta-2$, ~~$2\alpha\beta-2=N\alpha\beta+2\alpha$,~~ then
\begin{equation}
\label{24}
\alpha=\frac{n-N}{2-n}<-1,~~~~ \beta=\frac{2}{n-N}.
\end{equation}
Hence the system (\ref{14})-(\ref{16}) in the limit $t\to+\infty$ reduces to
\begin{equation}
\label{25}
\begin{array}{l}
0=K\left[ \frac{\alpha^2\beta^2}{2}{\phi_0}^2{(t-t_0)}^{2\alpha\beta-2}+V_0{\phi_0}^n{(t-t_0)}^{n\alpha\beta}-3\xi {H_0}^2{\phi_0}^N{(t-t_0)}^{N\alpha\beta+2\alpha}\right],
\end{array}
\end{equation}
\begin{equation}
\label{26}
\begin{array}{l}
0=-K\left[2\xi N H_0 \alpha\beta{\phi_0}^N{(t-t_0)}^{N\alpha\beta+\alpha-1}+2\xi \alpha H_0{\phi_0}^N{(t-t_0)}^{N\alpha\beta+\alpha-1}\right],
\end{array}
\end{equation}
\begin{equation}
\label{27}
\begin{array}{l}
\alpha\beta(\alpha\beta-1)\phi_0{(t-t_0)}^{\alpha\beta-2}+3\xi N{H_0}^2{\phi_0}^{N-1}{(t-t_0)}^{(N-1)\alpha\beta+2\alpha}+\\
+n V_0{\phi_0}^{n-1}{(t-t_0)}^{(n-1)\alpha\beta}=0.
\end{array}
\end{equation}
For
\begin{equation}
\label{28}
n=-N
\end{equation}
we have
\begin{equation}
\label{29}
\beta=-\frac{1}{N}, ~~~~\alpha=-{\frac{2 N}{N+2}}, ~~~~\alpha\beta=\frac{2}{N+2},
\end{equation}
and
\begin{equation}
\label{30}
{H_0}^2=\frac{2}{3\xi {(N+2)}^2}{\phi_0}^{2-N}+\frac{V_0}{3\xi{\phi_0}^{2 N}}.
\end{equation}
\\
\\
\\
\\
\text{~~~~}Therefore, summarizing the results of \textbf{(i)}, \textbf{(ii)} and neglecting $t_0$ in the limit $t\to+\infty$, we conclude that the following asymptotic solution 
\begin{equation}
\label{31}
H(t)=H_0{t}^{-{\frac{2 N}{N+2}}},
\qquad a(t)=a_0 e^{H_0\frac{N+2}{2-N}{t}^{\frac{2-N}{N+2}}},
\qquad \phi(t)=\phi_0{t}^{\frac{2}{N+2}}
\end{equation}
exists for $N>2$, ~~$n\leqslant-N$, ~~$t\to+\infty$.

    We note that the Hubble parameter decreases and tends to zero at the asymptotic solution (\ref{31}). The scale factor $a(t)$ increases slowly and approaches the constant $a_0$ in the limit $t\to+\infty$.

    The set of first-order differential equations (\ref{32}) is obtained from the initial equations (\ref{7})-(\ref{9}) for $\rho=0$ and is integrated numerically:
\begin{equation}
\begin{array}{l}
\label{32}
\dot{H}=-\frac{K\left( \Phi^2+2\xi H \Phi B'(\phi)\right)}{2(1+K\xi B(\phi))},\\
\dot\phi=\Phi,\\
\dot{\Phi}=-3 H\Phi-3\xi H^2 B'(\phi)-V'(\phi),\\
\dot a=a H,
\end{array}
\end{equation}
where the equality, following from (\ref{7}) for $\rho=0$,
$$
H^2=[\Phi^2+2 V(\phi)] /[6(1+K\xi B(\phi))]
$$
is checked at each step of numerical integration.

\begin{figure}[hbtp] 
\includegraphics[scale=0.44]{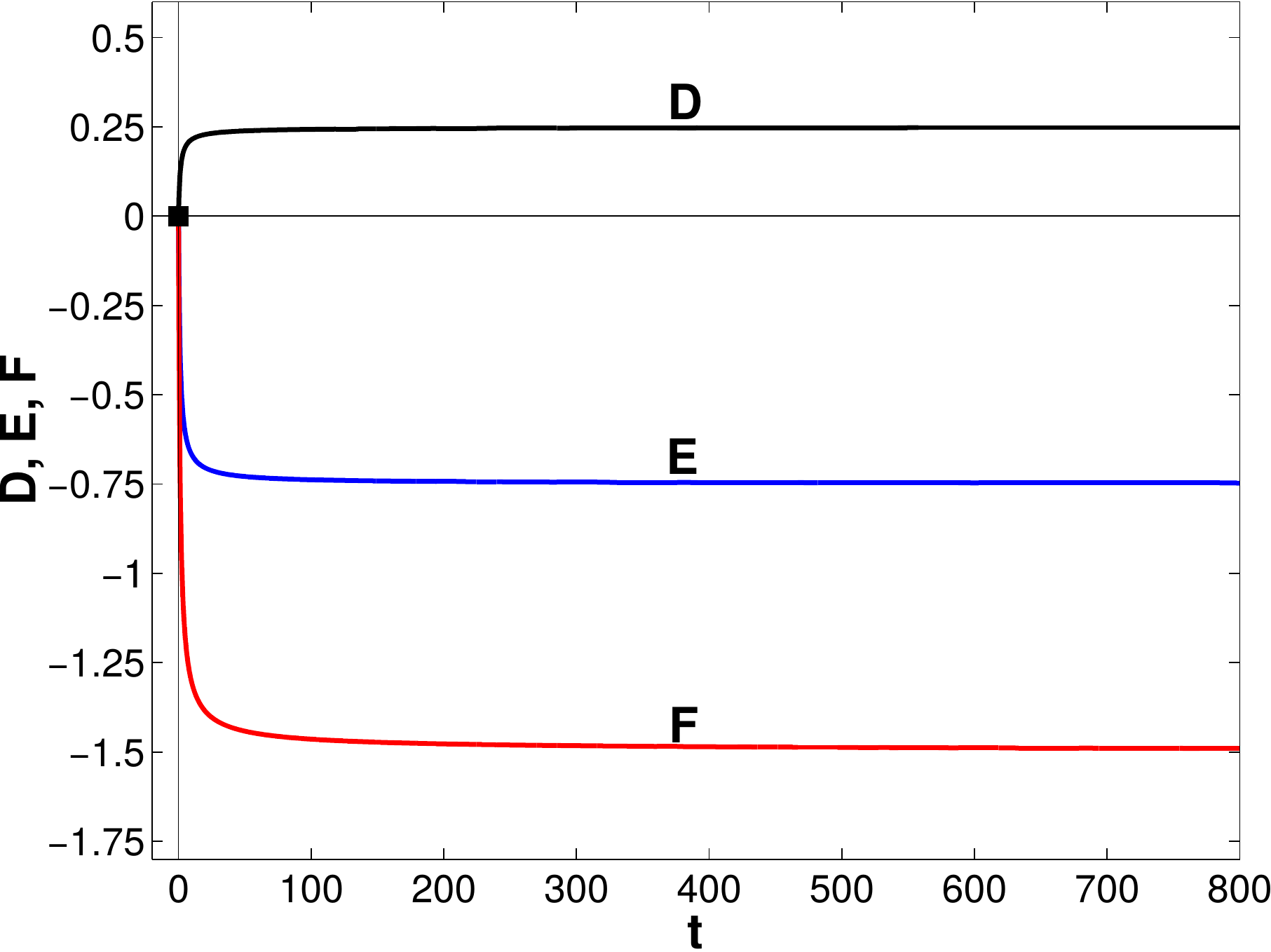}\qquad
\includegraphics[scale=0.44]{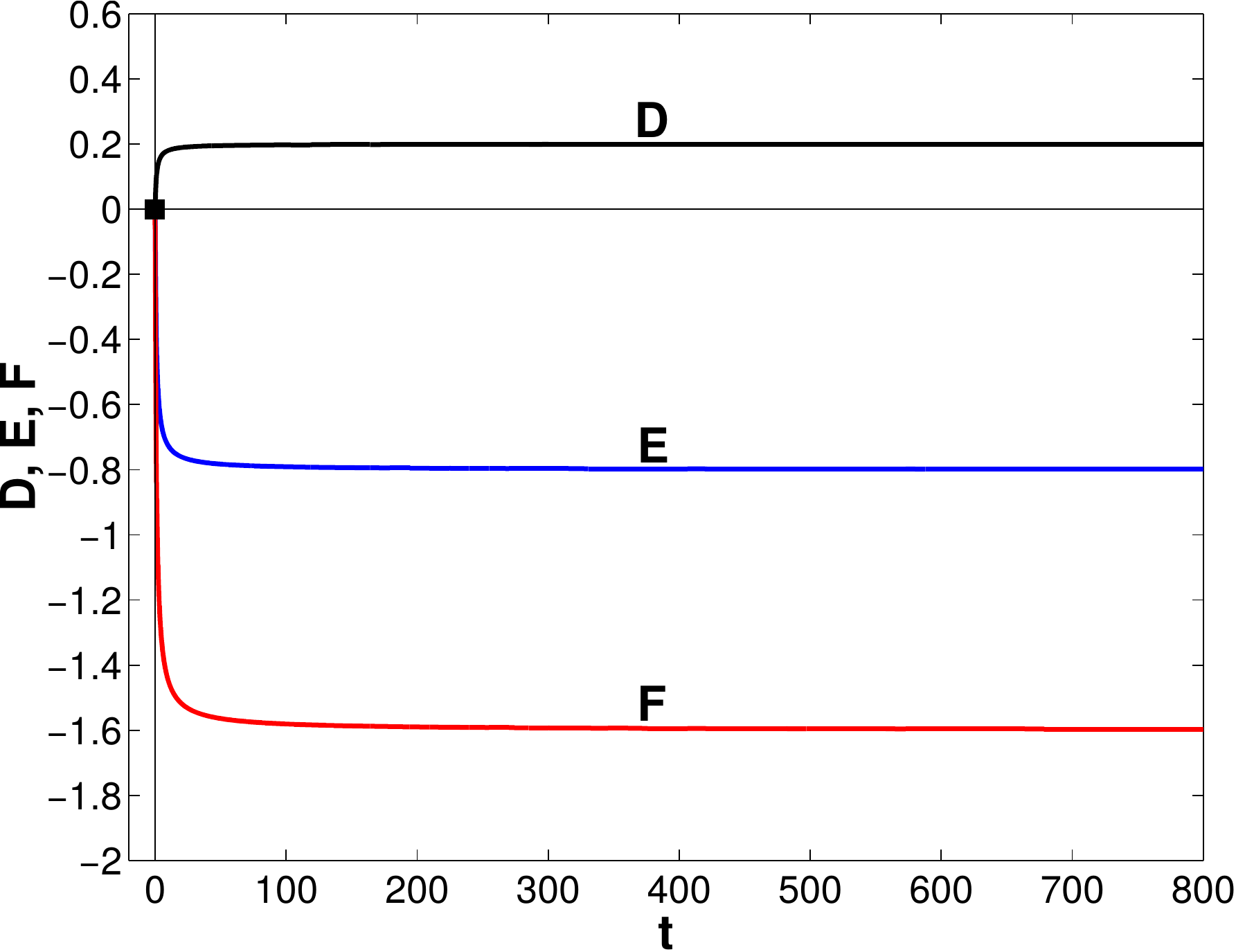}
\caption{Evolution of the quantities ~~$D(t)$, ~~$E(t)$, ~~$F(t)$~~ for the initial data: $\phi(0)=2$, ~~$\dot\phi(0)=0.4$, ~~$a(0)=1$. The parameters chosen: ~~$N=-n=6$~~ for the left plot and ~~$N=8$, ~~$n=-10$~~ for the right one. Other parameters: ~~$\xi=1$, ~~$V_0=1$, ~~$K=1$. Black squares are the initial values ~~$D(0)$, ~~$E(0)$, ~~$F(0)$.}
\label{Fig1}
\end{figure}

    We construct the following auxiliary quantities:
\begin{equation}
\label{33}
D(t)=\frac{d(\ln{(\phi)})}{d(\ln{(t)})}=\frac{t\dot\phi}{\phi}, ~~~~E(t)=\frac{d(\ln{(\dot\phi)})}{d(\ln{(t)})}=\frac{t\ddot\phi}{\dot\phi}, ~~~~F(t)=\frac{d(\ln{(H)})}{d(\ln{(t)})}=\frac{t\dot H}{H},
\end{equation} 
\\which are useful in the graphic representation since they approach constants if $\phi(t)$, $\dot\phi(t)$, $H(t)$ exhibit nearly a power-law behavior. Fig. \ref{Fig1} demonstrates the functions $D(t)$, $E(t)$, $F(t)$ obtained by numerical integration for fixed parameters $N\leqslant-n$. We see that ~~$D(t)\to\frac{2}{N+2}$, ~~$E(t)\to-\frac{N}{N+2}$, ~~$F(t)\to-\frac{2 N}{N+2}$~~ in the asymptotic regime (\ref{31}), $t\to+\infty$. Therefore, the power indices (\ref{29}) coincide with those found by numerical calculations.

    Moreover, the late-time tendency of auxiliary functions (\ref{33}) to the constants ~~$\frac{2}{N+2}$, ~~$\frac{N}{N+2}$, ~~$\frac{2 N}{N+2}$~~ is obtained numerically for a set of points of the plane $(\phi(0), \dot\phi(0))$ and for various values of the parameters $N$, $n$, $\xi$. Consequently, the obtained vacuum asymptotic solution is stable under homogeneous variations of the initial data. 

\begin{figure}[hbtp] 
\includegraphics[scale=0.44]{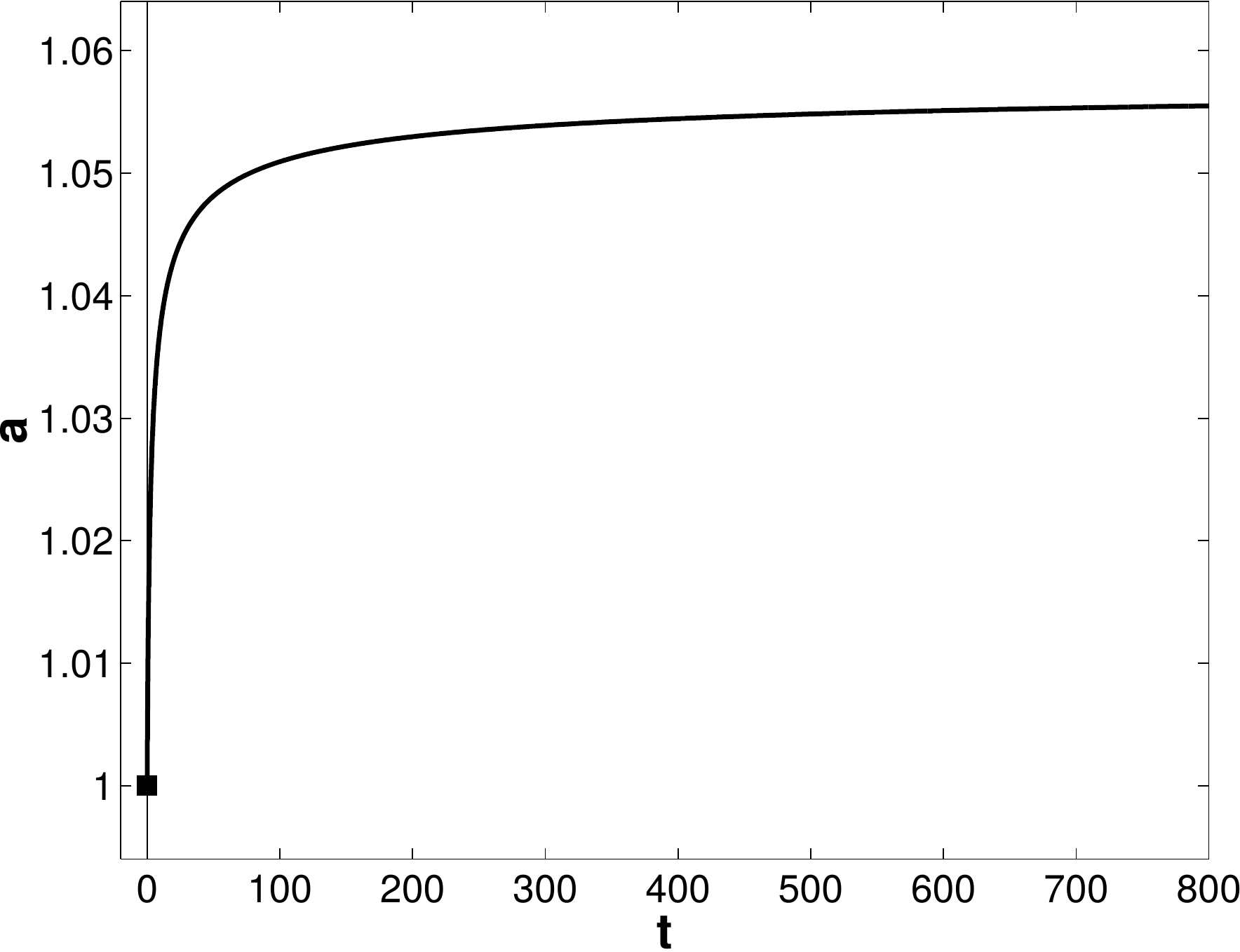}\qquad
\includegraphics[scale=0.44]{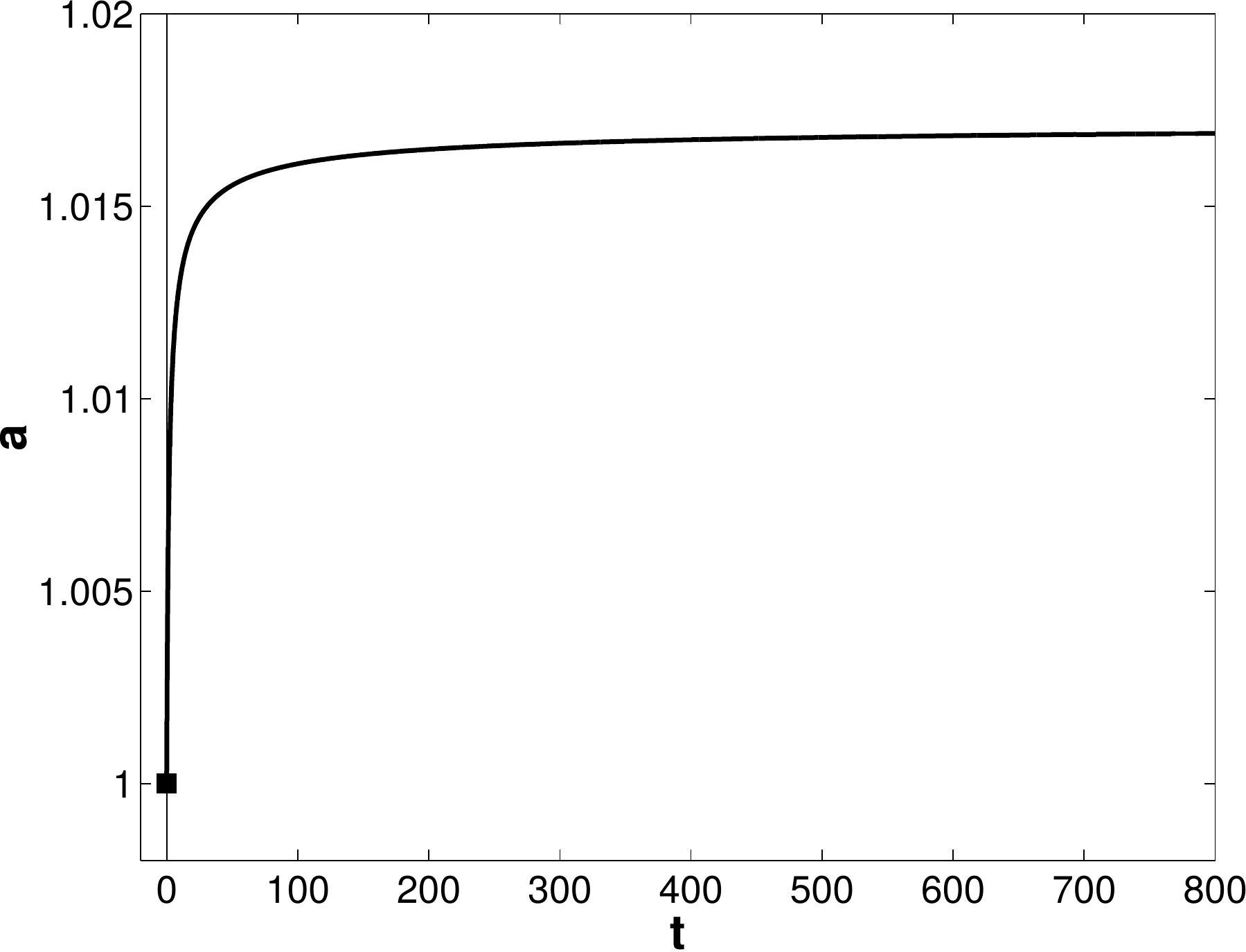}
\caption{Evolution of ~~$a(t)$~~ for the initial data $\phi(0)=2$, ~~$\dot\phi(0)=0.4$, ~~$a(0)=1$. Parameters: ~~$N=-n=6$~~ for the left graph and ~~$N=8$, ~~$n=-10$~~ for the right one. Other parameters: ~~$\xi=1$, ~~$V_0=1$, ~~$K=1$. Black squares are initial values ~~$a(0)$.}
\label{Fig2}
\end{figure}

\begin{figure}[hbtp] 
\includegraphics[scale=0.44]{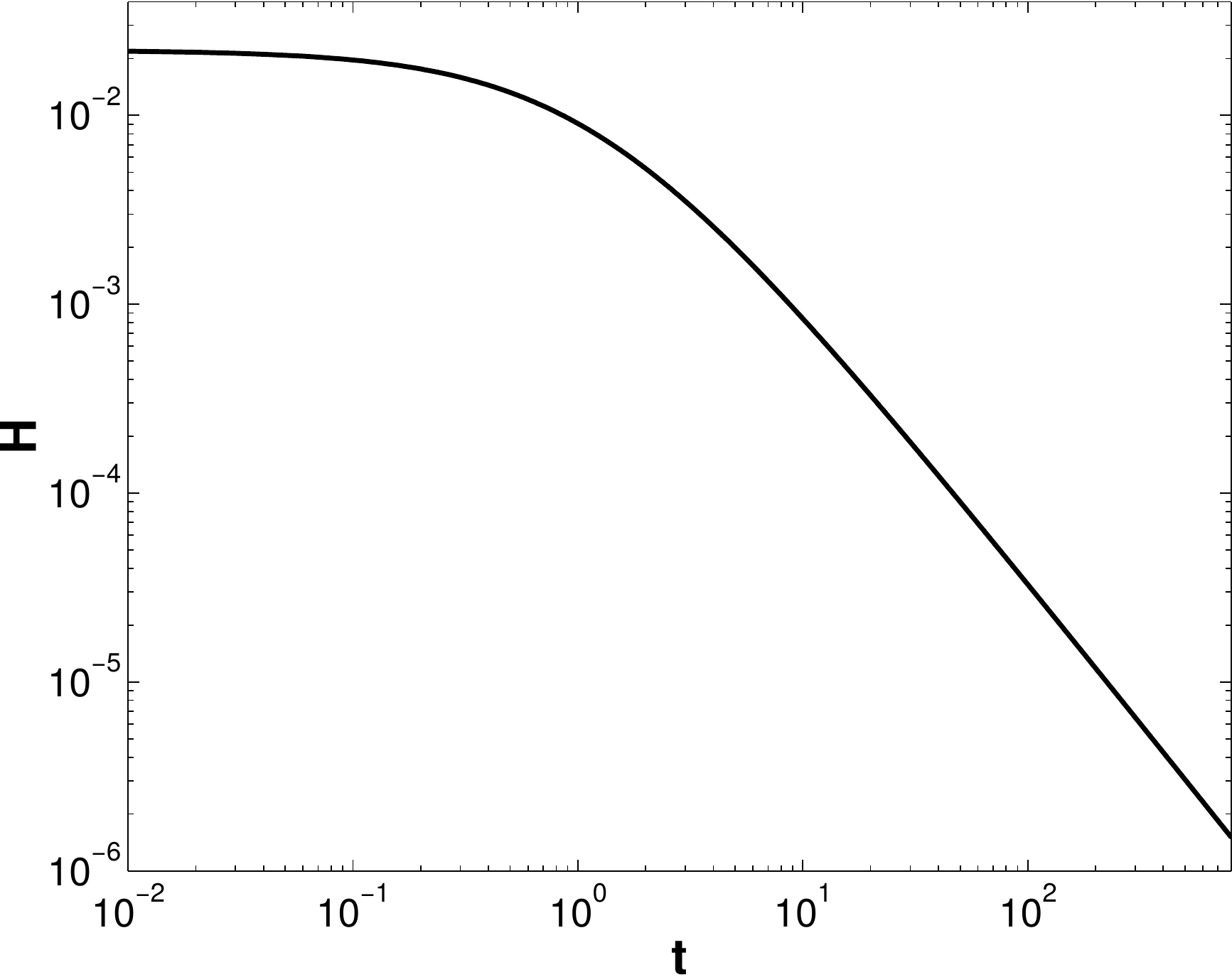}\qquad\quad
\includegraphics[scale=0.44]{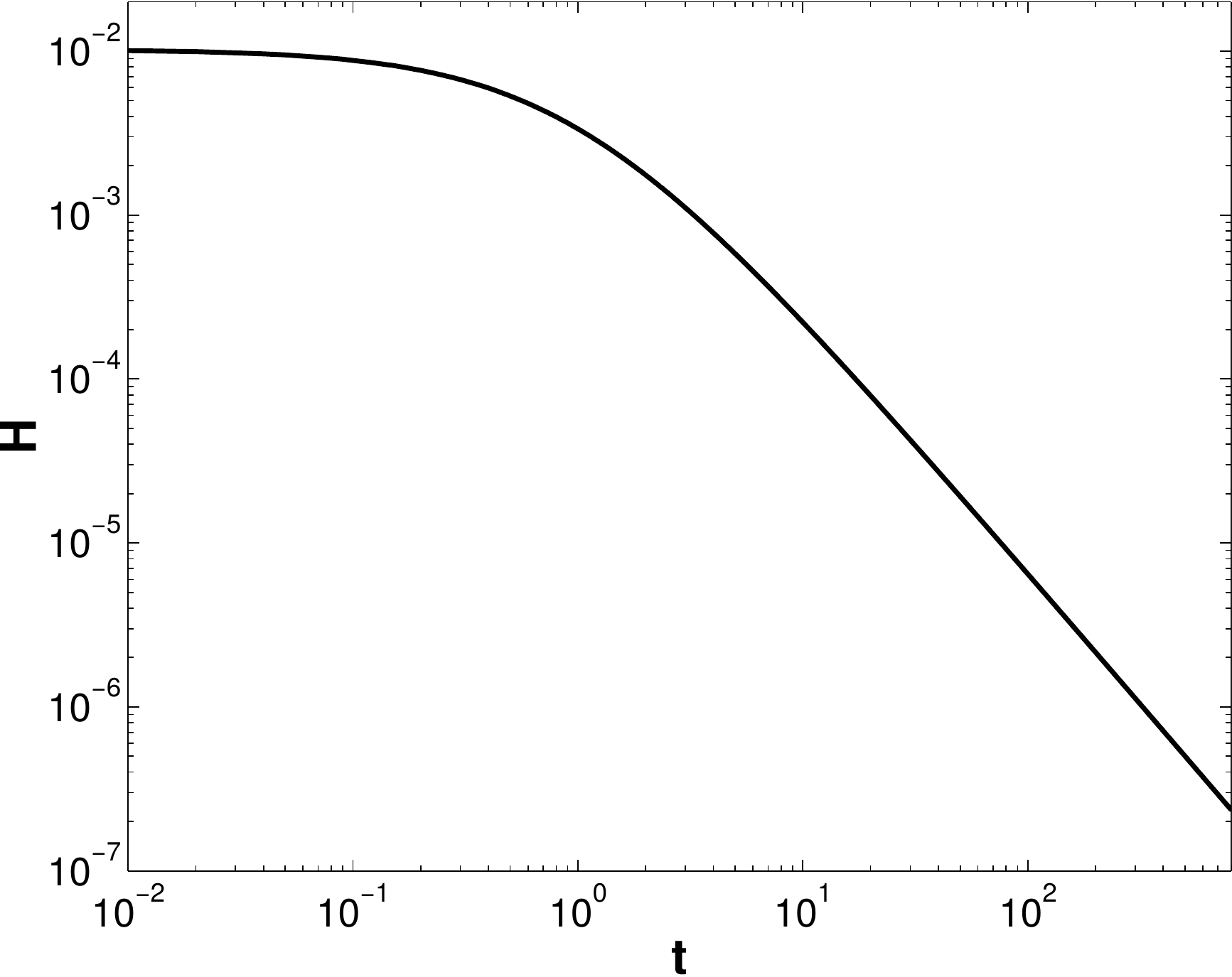}
\caption{Plots of ~~$H(t)$~~ are plotted for the initial data $\phi(0)=2$, ~~$\dot\phi(0)=0.4$, ~~$a(0)=1$. The starting value of the Hubble parameter $H(0)$ is calculated using the quadratic constraint equation (\ref{7}), we chose the positive root $H(0)>0$. Parameters: ~~$N=-n=6$~~ for the left graph and ~~$N=8$, ~~$n=-10$~~ for the right one. Other parameters: ~~$\xi=1$, ~~$V_0=1$, ~~$K=1$.}
\label{Fig3}
\end{figure}

\begin{figure}[hbtp] 
\includegraphics[scale=0.44]{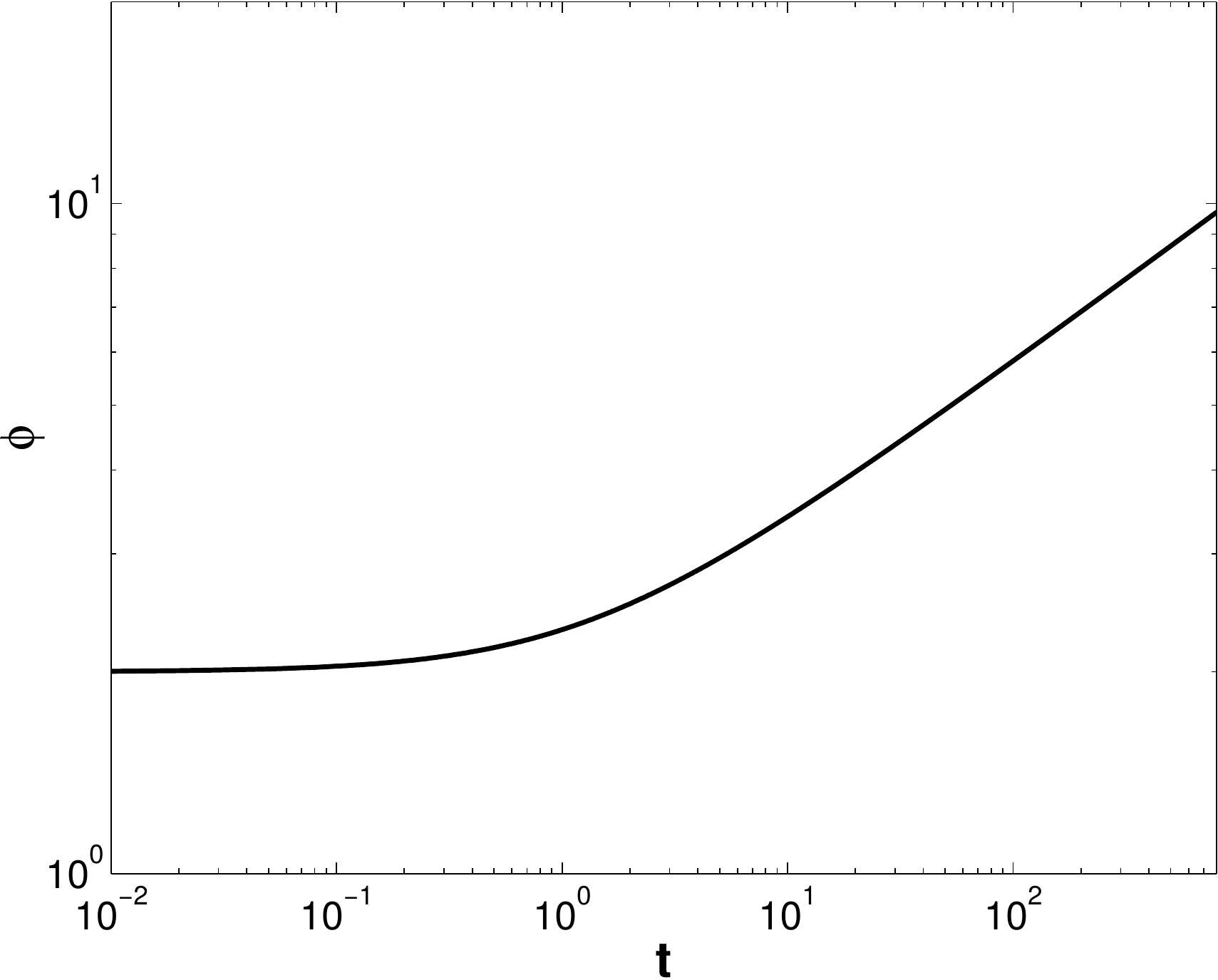}\qquad\quad
\includegraphics[scale=0.44]{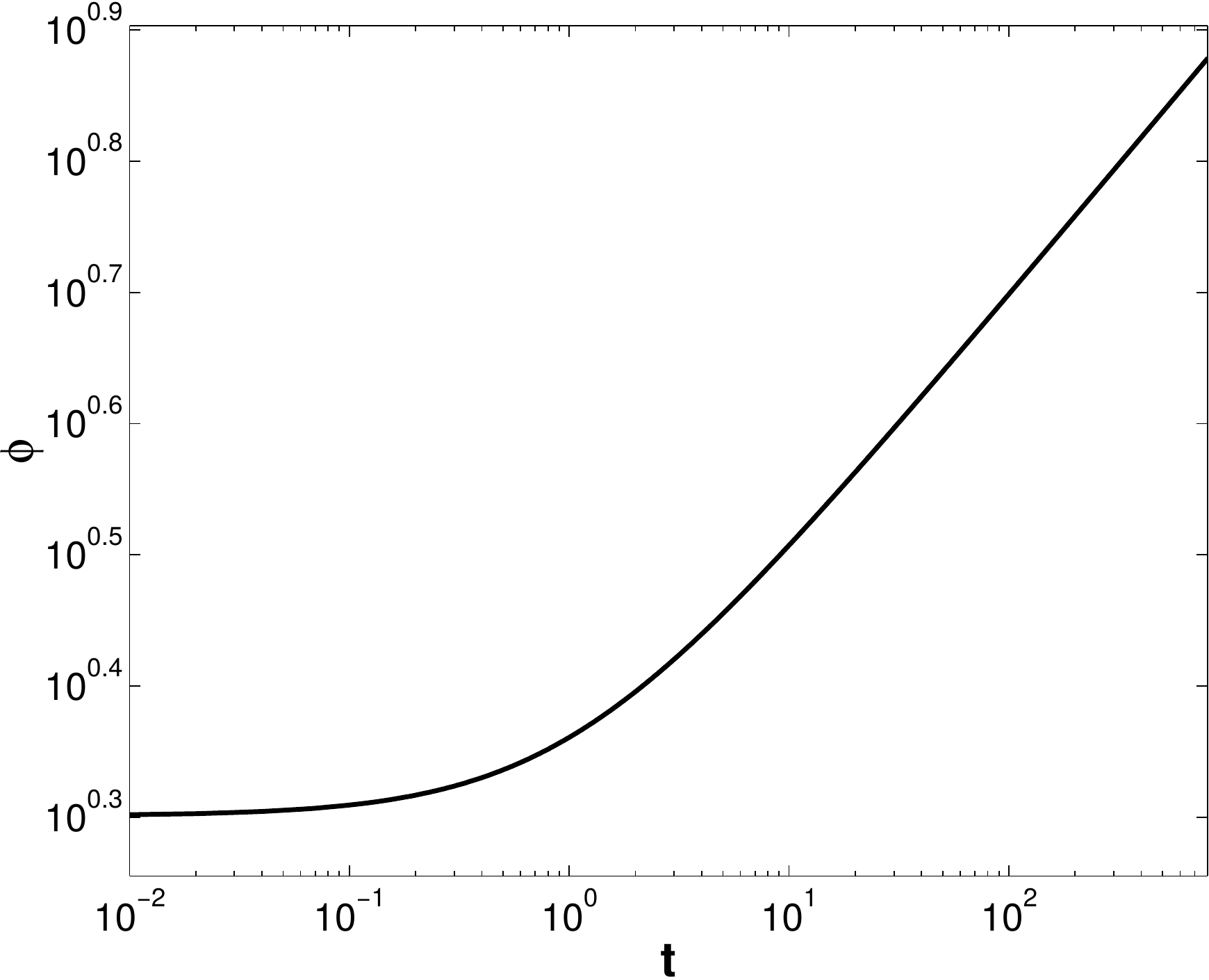}
\caption{Evolution of ~~$\phi(t)$~~ for the initial data $\phi(0)=2$, ~~$\dot\phi(0)=0.4$, ~~$a(0)=1$. Parameters: ~~$N=-n=6$~~ for the left plot and ~~$N=8$, ~~$n=-10$~~ for the right one. Other parameters: ~~$\xi=1$, ~~$V_0=1$, ~~$K=1$.}
\label{Fig4}
\end{figure}

    Dependences ~~$a(t)$, ~~$H(t)$, ~~$\phi(t)$~~ are plotted in Figs. \ref{Fig2}-\ref{Fig4} for the same initial data and parameters as in Fig. \ref{Fig1}.

\subsection{Models with Matter}
\label{III.2}
~~~~If matter is added ($\rho\neq 0$) to the model, then the power-law solution (\ref{31}) does not exist in the limit $t\to+\infty$. In this case, the matter term dominates compared with other terms in the equations of motion (\ref{7})-(\ref{8}). Namely, it follows from the continuity equation
\\$\dot\rho+3H(1+w)\rho=0$~~ that 
\begin{equation}
\label{34}
\rho(t)=\rho_0 e^{\frac{-3(1+w)H_0}{\alpha+1}{t}^{\alpha+1}}\to\rho_0,
\end{equation}
while 
\begin{equation}
\begin{array}{l}
\label{35}
\frac{{\dot\phi}^2}{2}=\frac{\alpha^2\beta^2}{2}{\phi_0}^2{t}^{2\alpha\beta-2}\to 0,\\
\\-3\xi H^2 \phi^N=-3\xi {H_0}^2{\phi_0}^N{t}^{N\alpha\beta+2\alpha}\to 0,\\
\\V(\phi)=V_0{\phi_0}^n{t}^{n\alpha\beta}\to 0
\end{array}
\end{equation}
for ~~$\alpha<-1$, ~~$\beta<0$, ~~$\alpha\beta<1$ ~~$N>2$, ~~$n\leqslant-N$,  ~~$t\to+\infty$.

    However, if the cosmological evolution begins from a very small value of $\rho(0)$, then the quasistatic phase with $a\approx const$ can be realized during some time interval. (This is the so-called ``loitering universe'', such a scenario in GR was described, for example, in \cite{Sahni}. A ``loitering universe'' is a transient stage of evolution with a very slow growth of the scale factor.) The quasistatic stage ends when the energy density of matter $\rho(t)$ begins to prevail over other terms in (\ref{7})-(\ref{8}).

\begin{figure}[hbtp] 
\includegraphics[scale=0.44]{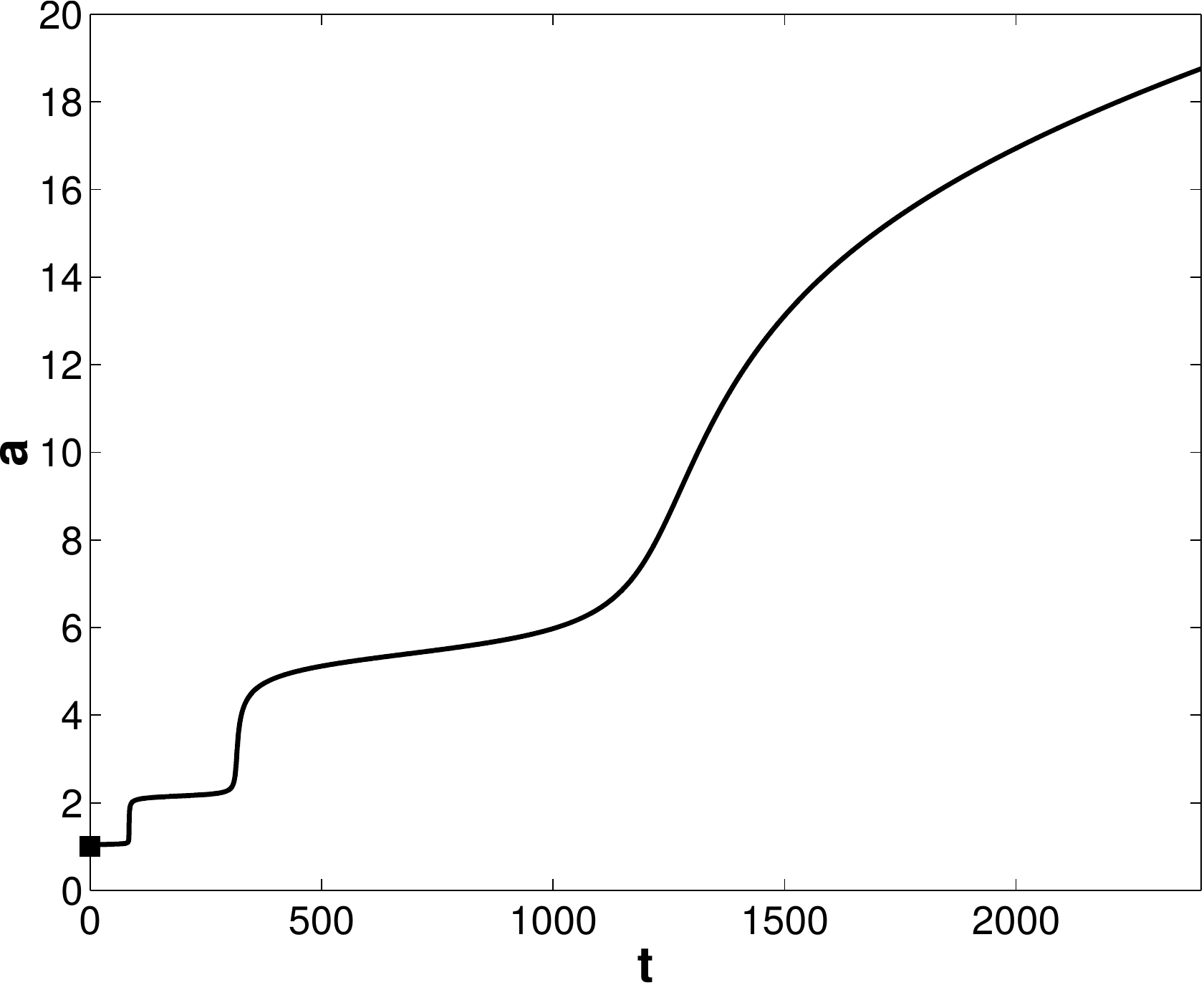}\qquad \quad
\includegraphics[scale=0.44]{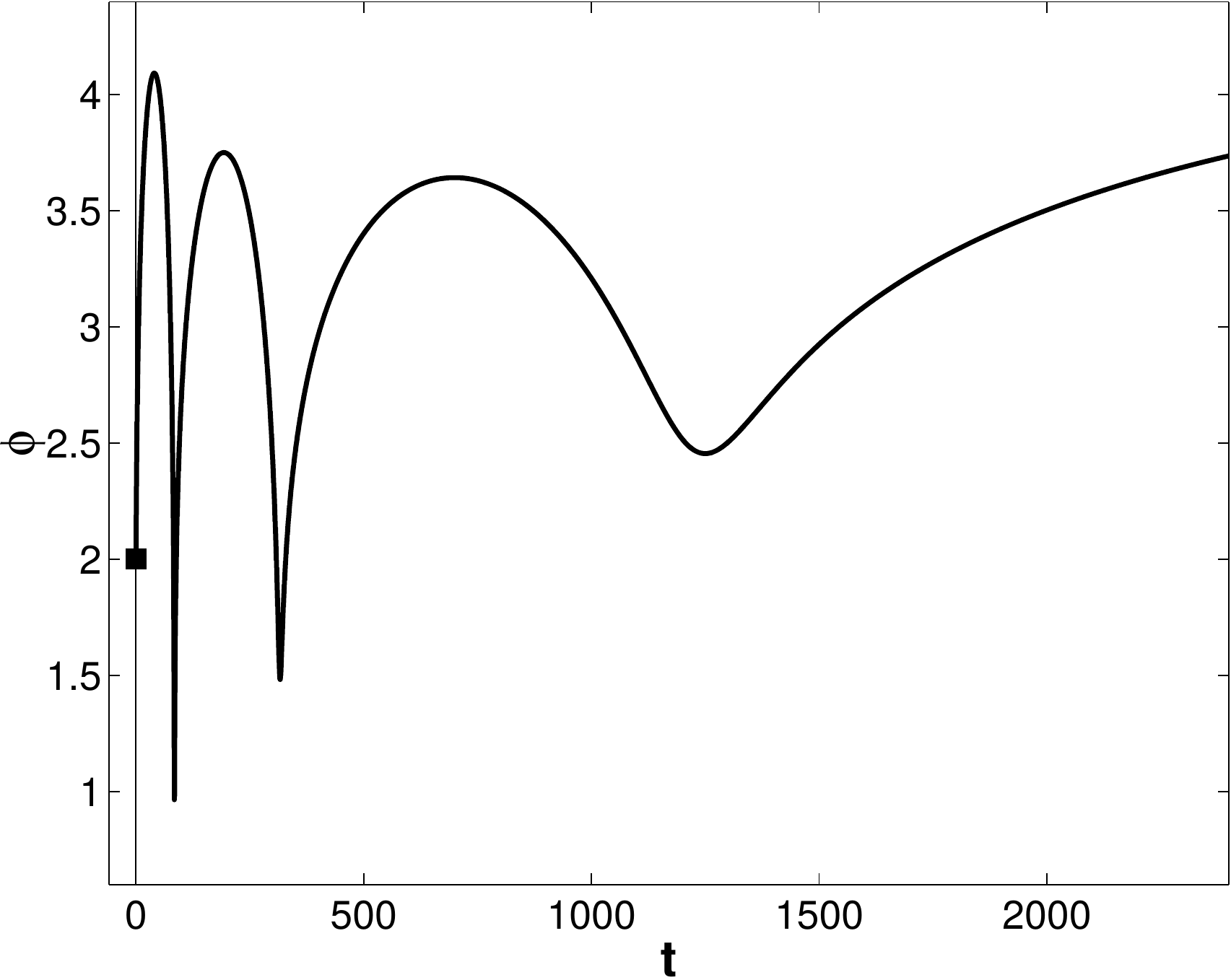}
\caption{Evolution of ~~$a(t)$, ~~$\phi(t)$~~ for the initial data: $\phi(0)=2$, ~~$\dot\phi(0)=0.4$, ~~$H(0)=0.02226$, ~~$a(0)=1$. Parameters: ~~$N=6$, ~~$n=-6$, ~~$\xi=1$, ~~$V_0=1$, ~~$w=0$, ~~$K=1$. Black squares are the initial values ~~$a(0)$, ~~$\phi(0)$.}
\label{Fig5}
\end{figure}

    Actually, the cosmological evolution in our models is more difficult to study for small $\rho(0)$. Using numerical integration of our set of equations,
\begin{equation}
\begin{array}{l}
\label{36}
\dot{H}=-\frac{K\left[ \Phi^2+2\xi H \Phi B'(\phi)+(1+w)\left(3 H^2(1/K+\xi B(\phi))-\Phi^2/2-V(\phi)\right)\right]  }{2(1+K\xi B(\phi)) },\\
\dot\phi=\Phi,\\
\dot{\Phi}=-3 H\Phi-3\xi H^2 B'(\phi)-V'(\phi),\\
\dot a=a H,
\end{array}
\end{equation}
where $\rho=3 H^2(1/K+\xi B(\phi))-\Phi^2/2-V(\phi)$ was substituted, we show that the scale factor behaves like the step function at early times for $\rho(0)\lesssim 10^{-3}$, and we have at least two temporary quasistatic stages (see Fig. \ref{Fig5}). It is worth emphasizing that the existence of more then one phases with $a\approx const$ is not obvious from the analytical form of the equations of motion, and they have been found only by numerical methods. Such a behavior of the scale factor is provided by scalar field oscillations.  

    Now ~~we ~check ~the ~existence ~of ~the ~asymptotic ~power-law ~regime ~in ~which ~~$a(t)=a_0{(t-t_0)}^{A}$, ~~$\left( H(t)=A{(t-t_0)}^{-1}\right)$, ~~$\phi(t)=\phi_0{(t-t_0)}^{AL}$, ~~$\rho(t)=\rho_0{(t-t_0)}^{-3(1+w)A}$, ~~where ~~$A$, ~~$L$, ~~$a_0$,~~${\phi}_0$,~~${\rho}_0$, ~~$t_0$~~ are constants. Substituting it to the initial system (\ref{7})-(\ref{9}) for ~~$B(\phi)=\phi^N$, ~~$V(\phi)=V_0\phi^n$,~~ we get
\begin{equation}
\label{37}
\begin{array}{l}
3A^2{(t-t_0)}^{-2}=K\left[ \frac{A^2 L^2}{2}{\phi_0}^2{(t-t_0)}^{2 A L-2}+V_0{\phi_0}^n{(t-t_0)}^{n AL}-\right. \\
~~~~~~~~~~~~~~~~~~~~\left. -3\xi A^2{(t-t_0)}^{-2}{\phi_0}^N{(t-t_0)}^{N AL}+\rho_0{(t-t_0)}^{-3(1+w)A}\right],
\end{array}
\end{equation}
\begin{equation}
\label{38}
\begin{array}{l}
-2 A{(t-t_0)}^{-2}=-K\left[ A^2 L^2{\phi_0}^2{(t-t_0)}^{2 A L-2}+2\xi N A {(t-t_0)}^{-1} AL{\phi_0}^N{(t-t_0)}^{N AL-1}-\right. \\
~~~~~~~~~~~~~~~~~~~~~~\left. -2\xi A {(t-t_0)}^{-2}{\phi_0}^N{(t-t_0)}^{N AL}+(1+w)\rho_0{(t-t_0)}^{-3(1+w)A}\right] ,
\end{array}
\end{equation}
\begin{equation}
\label{39}
\begin{array}{l}
A L(A L -1)\phi_0{(t-t_0)}^{A L-2}+3 A{(t-t_0)}^{-1} A L\phi_0{(t-t_0)}^{AL-1}+\\
+3\xi N A^2{(t-t_0)}^{-2} {\phi_0}^{N-1}{(t-t_0)}^{(N-1)A L}+n V_0{\phi_0}^{n-1}{(t-t_0)}^{(n-1)AL}=0.
\end{array}
\end{equation}
If ~~$A>0$, ~~$L>0$, ~~$N>2$, ~~$n<0$,~~ and ~~$n A L=N A L-2$,~~ $N A L -2=-3(1+w)A$,~~ then
\begin{equation}
\label{40}
A=\frac{2n}{3(1+w)(n-N)},~~~~ L=-\frac{3(1+w)}{n}.
\end{equation}
Keeping only dominant terms in the limit $t\to+\infty$, from Eqs.~(\ref{37})-(\ref{39}) it follows
\begin{equation}
\label{41}
\begin{array}{l}
0=K\left[ V_0{\phi_0}^n{(t-t_0)}^{n AL}-3\xi A^2{\phi_0}^N{(t-t_0)}^{N AL-2}+\rho_0{(t-t_0)}^{-3(1+w)A}\right] ,
\end{array}
\end{equation}
\begin{equation}
\label{42}
\begin{array}{l}
0=-K\left[2\xi N A^2 L{\phi_0}^N{(t-t_0)}^{N A L-2}-2\xi A {\phi_0}^N{(t-t_0)}^{N AL-2}+\right. \\
~~~~~\left. +\rho_0(1+w){(t-t_0)}^{-3(1+w)A}\right] ,
\end{array}
\end{equation}
\begin{equation}
\label{43}
\begin{array}{l}
3\xi N A^2{\phi_0}^{N-1}{(t-t_0)}^{(N-1)A L-2}+n V_0{\phi_0}^{n-1}{(t-t_0)}^{(n-1)AL}=0
\end{array}
\end{equation}
and we find
\begin{equation}
\begin{array}{l}
\label{44}
\rho_0=\frac{4 n\xi{\phi_0}^N(N+n)}{3{(N-n)}^2{(1+w)}^2}>0 \quad\Rightarrow \quad n<-N,\\
V_0=-\frac{4 n\xi N{\phi_0}^{N-n}}{3{(N-n)}^2{(1+w)}^2}.
\end{array}
\end{equation}
We obtain the asymptotic solution, neglecting $t_0$ for $t\to+\infty$:
\begin{equation}
\label{45}
a(t)=a_0{t}^{{\frac{2 n}{3(1+w)(n-N)}}},
\qquad \phi(t)=\phi_0 {t}^{\frac{2}{N-n}},
\qquad \rho(t)=\rho_0{t}^{\frac{2n}{N-n}},
\end{equation}
which exists for $N>2$, ~~$n<-N$, ~~$t\to+\infty$.

    The scale factor and the scalar field increase in the asymptotic solution (\ref{45}), while the Hubble parameter and the energy density of matter decrease to zero as $t\to+\infty$. 

    Taking ~~into ~~account ~~(\ref{10}), ~(\ref{11}), ~we ~obtain ~from ~Eqs.~(\ref{41}) ~and ~(\ref{42}) ~the ~functions
\\$\rho_{\phi}(t)=-\rho(t)$, ~~$p_{\phi}(t)=-p(t)$~~ and therefore $w_{\phi}=\frac{p_{\phi}}{\rho_{\phi}}=w=\frac{p}{\rho}$. This is the property of a scaling solution. An analogous regime has been found earlier for $N=2$ in \cite{Masha2}. (The scaling solution in models with the scalar field and matter is a solution in which the equation-of-state parameter of a scalar field equals is the same as for matter, see, e.g., \cite{Amendola},~p.~140,~~ \cite{Wands}).

\begin{figure}[hbtp] 
\includegraphics[scale=0.44]{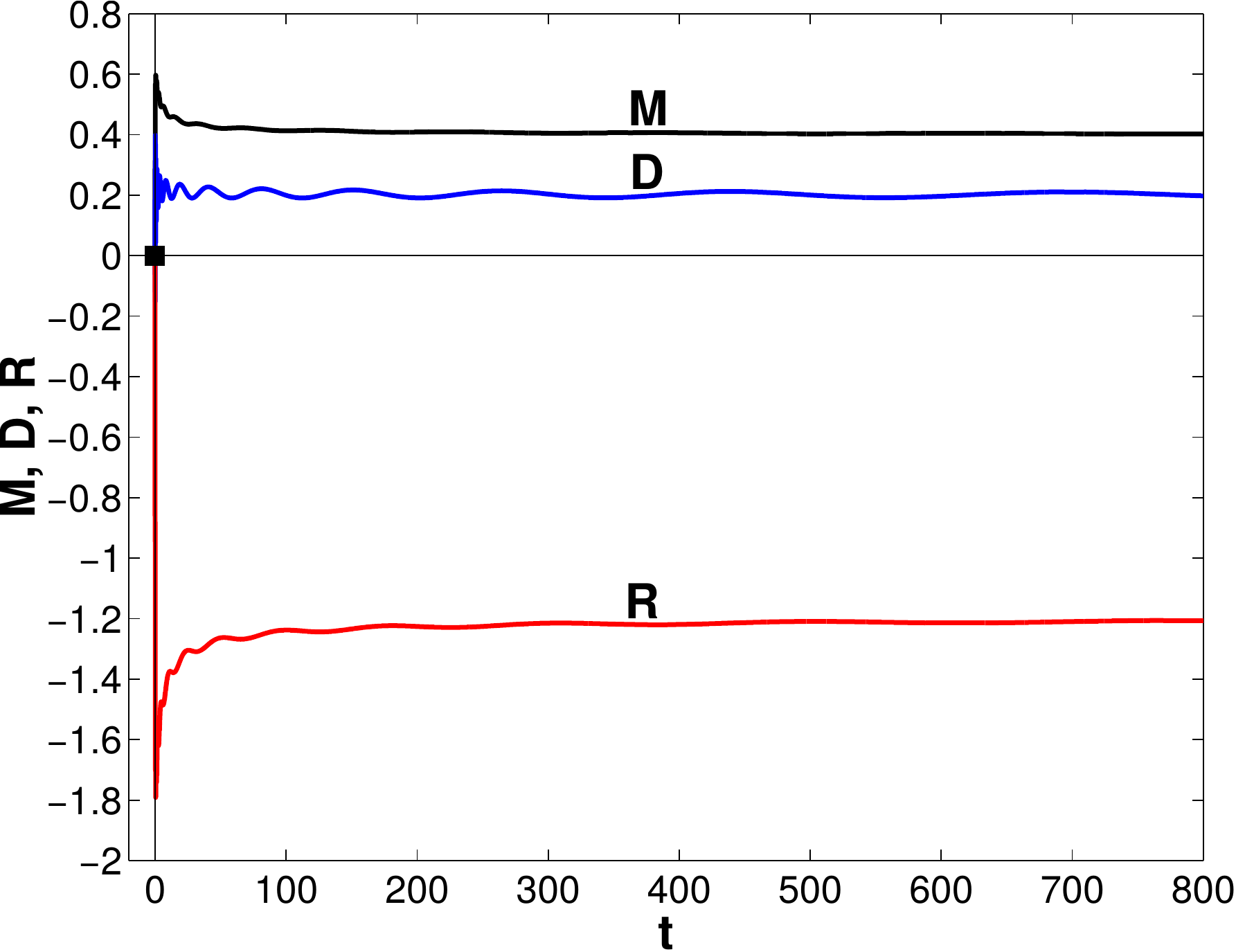}\qquad \quad
\includegraphics[scale=0.44]{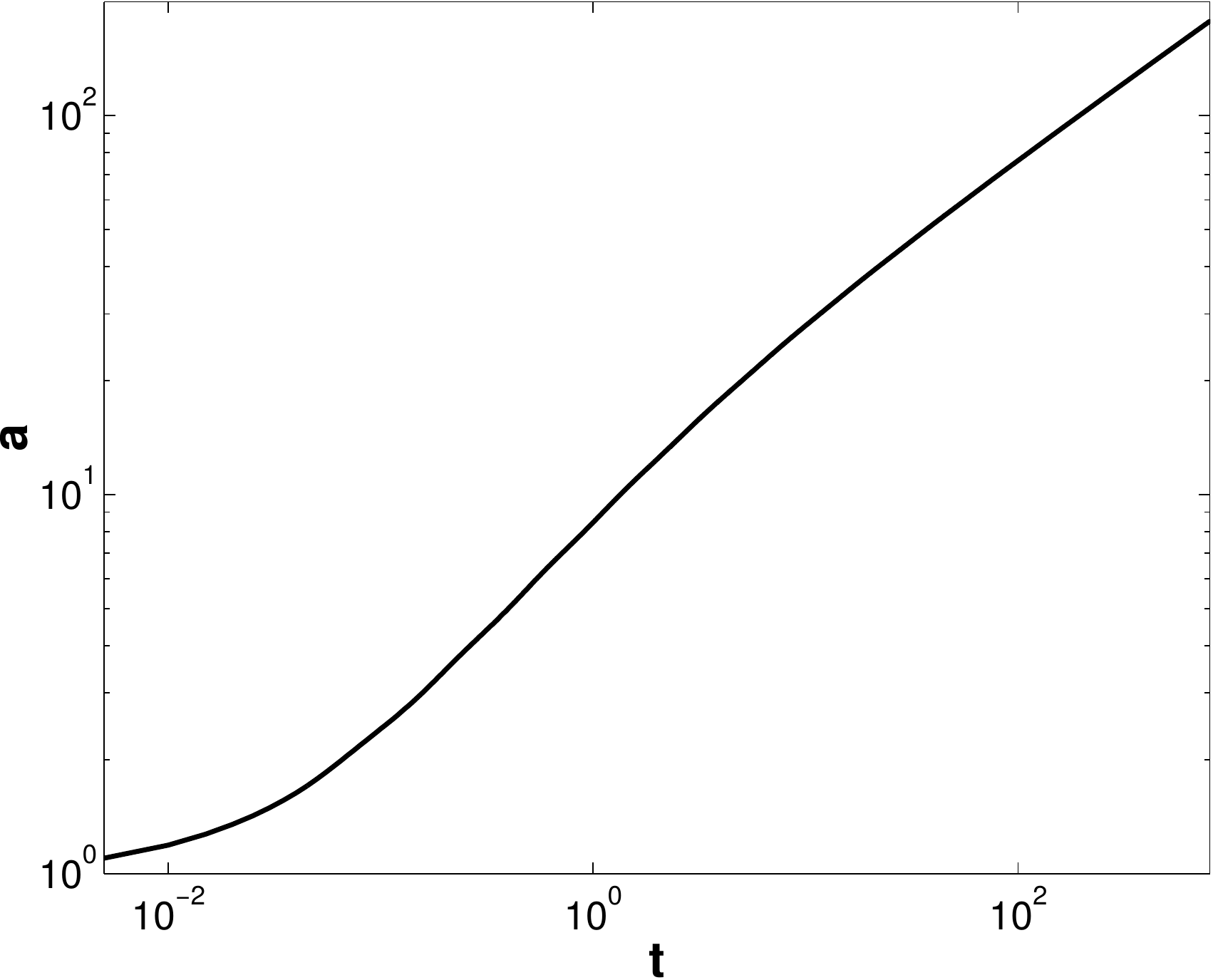}
\caption{Evolution of the quantities ~~$M(t)$, ~~$D(t)$, ~~$R(t)$~~ in the left graph and of the scale factor $a(t)$ in the right one for the initial data $\phi(0)=0.4$, ~~$\dot\phi(0)=2$, ~~$H(0)=20$, ~~$a(0)=1$. Parameters: ~~$N=4$, ~~$n=-6$, ~~$\xi=4$, ~~$V_0=1$, ~~$w=0$, ~~$K=1$. Black squares mark the initial values ~~$M(0)$, ~~$D(0)$, ~~$R(0)$. }
\label{Fig6}
\end{figure}

    Auxiliary functions are introduced in the same way as in the previous subsection:
\begin{equation}
\label{46}
M(t)=\frac{d(\ln{(a)})}{d(\ln{(t)})}=t H, ~~~~D(t)=\frac{d(\ln{(\phi)})}{d(\ln{(t)})}=\frac{t\dot\phi}{\phi}, ~~~~R(t)=\frac{d(\ln{(\rho)})}{d(\ln{(t)})}=-3(1+w)t H.
\end{equation} 
These quantities have the simple form of a constant in corresponding plots when the cosmological evolution passes through the power-law stage. The time dependences of these functions are plotted in Fig. \ref{Fig6} (left). These functions approach the power indices ~~$M(t)\to\frac{2n}{3(1+w)(n-N)}$, ~~$D(t)\to\frac{2}{N-n}$, ~~$R(t)\to\frac{2n}{N-n}$~~ in the asymptotic regime (\ref{45}), $t\to+\infty$.

    Numerical integration has been carried out for the set points of space $(\phi(0), \dot\phi(0), H(0))$ and for various parameters $N$, $n$, $\xi$. We have found that the quantities $M(t)$, $D(t)$, $R(t)$ tend to constants $\frac{2n}{3(1+w)(n-N)}$, $\frac{2}{N-n}$, $\frac{2n}{N-n}$, respectively. Therefore, the power-law solution (\ref{45}) is stable with under homogeneous variations of initial data. 
 
\begin{figure}[hbtp] 
\includegraphics[scale=0.44]{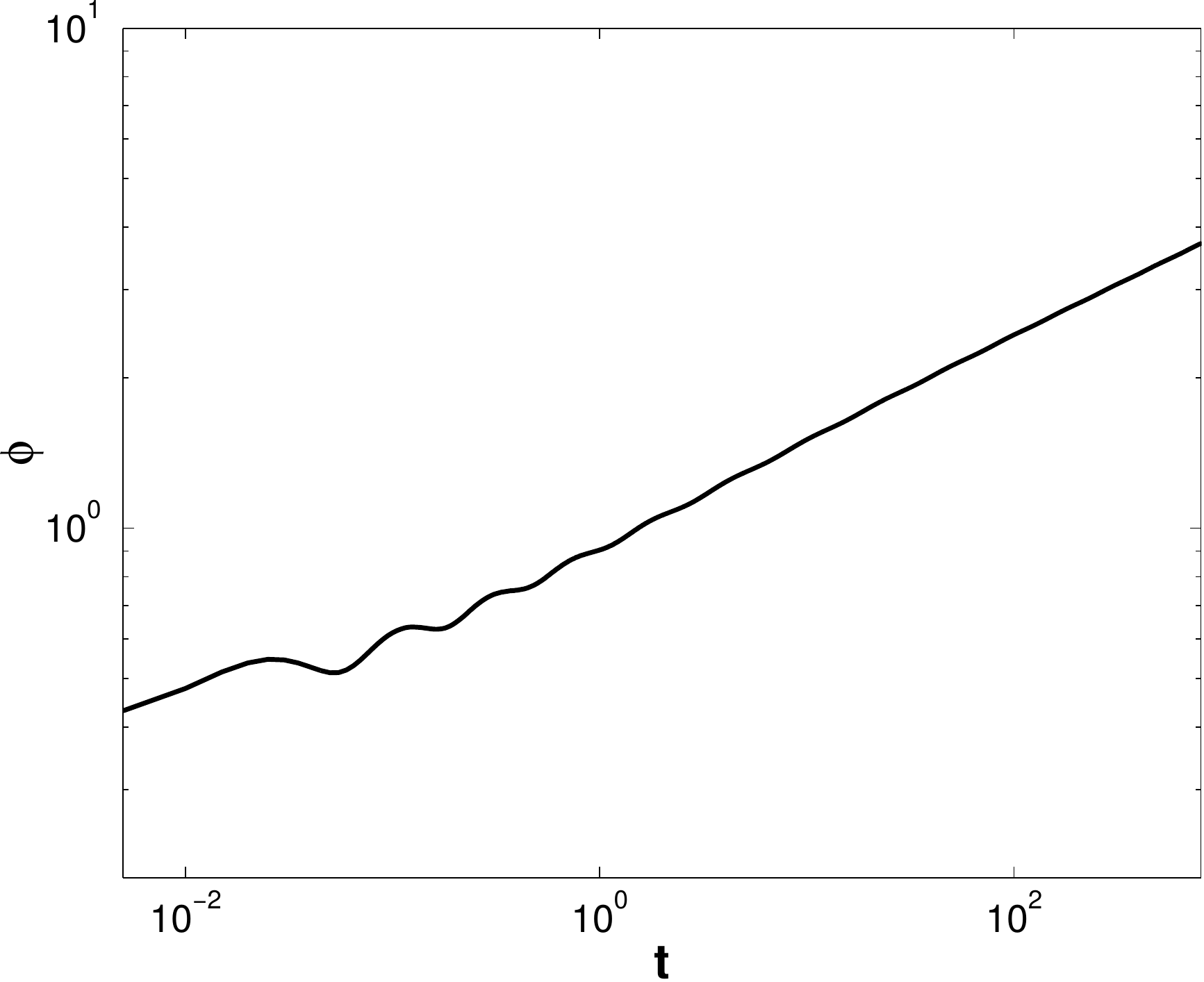}\qquad \quad
\includegraphics[scale=0.44]{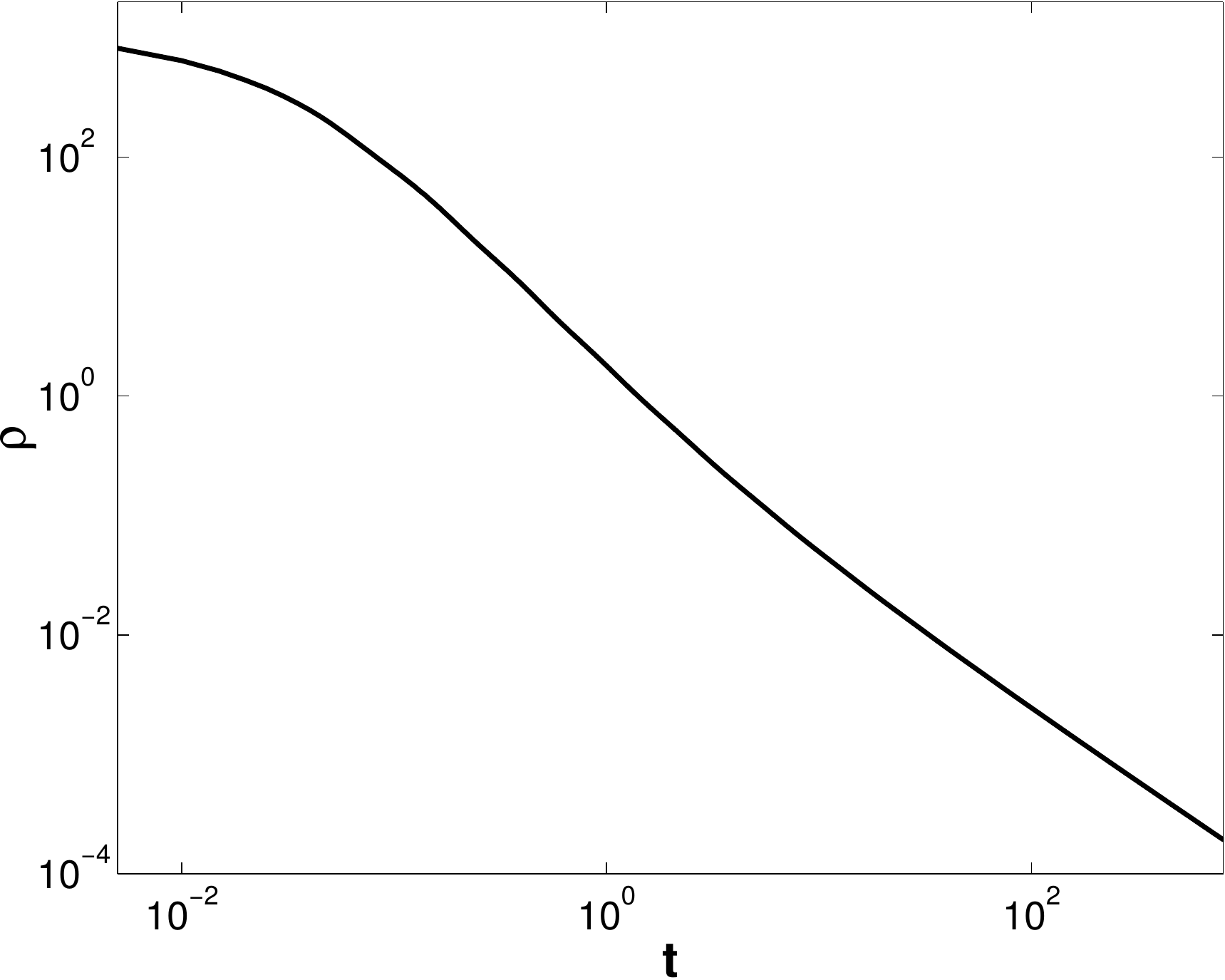}
\caption{Evolution ~~of ~~$\phi(t)$, ~~$\rho(t)$~~ for ~~the ~~initial ~~data ~~$\phi(0)=0.4$, ~~$\dot\phi(0)=2$, ~~$H(0)=20$, ~~$a(0)=1$. The starting value of the energy density $\rho(0)$ of matter is calculated using the constraint equation (\ref{7}). Parameters: ~~$N=4$, ~~$n=-6$, ~~$\xi=4$, ~~$V_0=1$, ~~$w=0$, ~~$K=1$.}
\label{Fig7}
\end{figure}

    The behavior of the scale factor, the scalar field and the energy density of matter are shown in Fig. \ref{Fig6} (right), Fig. \ref{Fig7}, where the initial conditions and parameters are the same as in Fig. \ref{Fig6}.

    Finally, we note that asymptotic power-law solutions have not been revealed neither numerically nor analytically in the matter case for $N=-n$. This case needs a further analysis.

\section{Conclusion}
\label{IV}
~~~~In the present paper the evolution of the late Universe has been studied in models of teleparallel gravity with nonminimal coupling ~$\xi T\phi^N$, ~$N>2$, ~$\xi>0$~, the field potential ~$V(\phi)=V_0\phi^n$, ~$n<0$ and a perfect fluid with the equation of state $p=w\rho$. We find analytically new asymptotic solutions in the vacuum case and in the presence of a perfect fluid for $t\to+\infty$. 

    The vacuum asymptotic regime $H(t)=H_0 {t}^{-\frac{2N}{N+2}}\to 0$ ~~($a(t)\to~const$), ~~$\phi(t)=\phi_0{t}^{\frac{2}{N+2}}\to\infty$ exists only for $N>2$, $n\leqslant-N$. The found asymptotic solution exists when the term with the nonminimal coupling, the kinetic energy of the scalar field for $n<-N$ and the potential for $n=-N$ dominate. We note that the deceleration parameter $q\equiv-\frac{\ddot a}{aH^2}$ corresponding to the obtained solution is 
$$q=-1+\frac{2N}{H_0(N+2)}{t}^{\frac{N-2}{N+2}}\to+\infty, ~~t\to+\infty.
$$
Therefore, the Universe expands with deceleration at late times and approaches a static state. 
    
    We have shown numerically (see Fig.~\ref{Fig1}-\ref{Fig4}) that our vacuum solution is stable with respect to homogeneous variations of the initial data. Therefore, Einstein's plan to receive a stable static Universe in GR is realized in the model of teleparallel gravity with a nonminimally coupled scalar field. This result is rather surprising since not so many stable stationary cosmological models are known.
    
    Any perfect fluid ($\rho\neq 0$) destroys the vacuum asymptotic solution for ~~$t\to+\infty$~~ since in this case the energy density of matter prevails over other components of the field equations. However, if the cosmological evolution starts from a small $\rho$, then it passes though several transient quasistatic stages with $a\approx const$ similarly to a ``loitering universe'' in GR.

    In models with matter, other stable asymptotic regime exists for $N>2$, $n<-N$: 
$$
a(t)=a_0 {t}^{\frac{2n}{3(1+w)(n-N)}}\to +\infty, ~~\phi(t)=\phi_0{t}^{\frac{2}{N-n}}\to\infty, ~~\rho(t)=\rho_0{t}^{\frac{2n}{N-n}}\to 0, ~~t\to+\infty.
$$
The corresponding deceleration parameter is 
$$
q=-1+\frac{3(1+w)(n-N)}{2n}>-1.
$$ 
Therefore, an accelerated expansion is possible for $w\leqslant-\frac{2}{3}$. The nonminimal coupled term, the potential energy and the matter energy density dominate at this asymptotic solution. It is a scaling one as $w_{\phi}=w$. Such a behavior of the scalar field can be suitable for a description of radiation- and matter-dominated stages of the Universe.

    Numerical integration confirms the existence of the scaling solution for $t\to+\infty$, which is stable with respect to homogeneous variations of the initial data. Plots in Fig.~\ref{Fig6},~\ref{Fig7} (for fixed $N$ and $n$) show that the functions $a(t)$, $\phi(t)$ and $\rho(t)$ approach a power-law behavior and the corresponding powers coincide with those found analytically.  

    We write down a possible final of the cosmological evolution in our models with ~~$N>2$, ~~$n<0$:
\\
\\\textbf{1.} $\rho=0$.\\
\\\text{~~~~}\textbf{(i)} $-N<n<0$ --- stable de Sitter solution (see \cite{Masha1}, \cite{Masha2}) $H=H_0$, ~~$\phi=\phi_0$, ~~$q=-1$,\\
\\\text{~~~~}\textbf{(ii)} $n\leqslant-N$ --- stable asymptotic solution:  
~~$H(t)=H_0{t}^{-\frac{2N}{N+2}}$, ~~$\phi=\phi_0{t}^{\frac{2}{N+2}}$,
\\\text{~~~~~~~~~~}$q=-1+\frac{2N}{H_0(N+2)}{t}^{\frac{N-2}{N+2}}\to+\infty$, ~~$t\to+\infty$.\\
\\
\\
\\\textbf{2.} $\rho\neq 0$.\\
\\\text{~~~~}\textbf{(i)} $-N<n<0$ --- stable de Sitter solution (see \cite{Masha2}) $H=H_0$, ~~$\phi=\phi_0$, ~~$q=-1$,\\
\\\text{~~~~}\textbf{(ii)} $n<-N$ --- stable asymptotic solution: 
~~$a(t)=a_0{{t}^{\frac{2n}{3(1+w)(n-N)}}}$, ~~$\phi=\phi_0{t}^{\frac{2}{N-n}}$, ~~$\rho=\rho_0{t}^{\frac{2n}{N-n}}$,
\\\text{~~~~~~~~~~}$q=-1+\frac{3(1+w)(n-N)}{2n}=const>-1$, ~~$t\to+\infty$.\\
\\
\\

    Therefore, we see that strongly decreasing potentials ($n\leqslant-N$) lead to a late-time behavior of cosmological quantities different from the one at the acceleration stage of the real Universe. However, models with a decreasing potential and a perfect fluid might be interesting due to the existence of a scaling solution, which can be used for constructing realistic cosmological models.

\section*{Acknowledgements}
~~~~The author is grateful to Alexey Toporensky for useful discussions. This work was supported by RSF Grant \textnumero 16-12-10401.

\end{document}